\documentclass[12pt]{iopart}

\usepackage{iopams}
\usepackage{graphicx}
\usepackage{amssymb}
\usepackage{color}
\usepackage{ifthen}
\usepackage{amsfonts}

\newcommand{\tk}{\tilde{\kappa}}

\begin{document}

\title{On Numerical relativistic hydrodynamics and barotropic equations of state} 

\author{Jos\'e Mar\'ia Ib\'a\~nez$^1$, Isabel Cordero-Carri\'on$^2$ and Juan Antonio Miralles$^3$}

\address{$^1$ Department of Astronomy and Astrophysics, University of Valencia,
E-46100, Burjassot, Spain}
\address{$^2$ Max-Planck-Institute for Astrophysics, Garching, 
Karl-Schwarzschild-Str. 1, D-85741, Garching, Germany}
\address{$^3$ Department of Applied Physics, University of Alicante,
Campus de Sant Vicent del Raspeig, E-03080, Alicante, Spain
}
\ead{jose.m.ibanez@uv.es, chabela@mpa-garching.mpg.de, ja.miralles@ua.es}

\begin{abstract}
The characteristic formulation of the relativistic hydrodynamic equations 
(Donat {\it et al} 1998 {\it J. Comput. Phys.} {\bf 146} 58), which has been 
implemented in many relativistic hydro-codes that make use of Godunov-type 
methods, has to be slightly modified in the case of evolving barotropic flows. 
For a barotropic equation of state, a removable singularity appears in one of 
the eigenvectors. The singularity can be avoided by means of a simple 
renormalization which makes the system of eigenvectors well defined and 
complete. An alternative strategy for the particular case of barotropic flows 
is discussed.
\end{abstract}

\pacs{04.25.D-, 47.11.-j, 47.75.+f, 95.30.Sf}

\maketitle

Barotropic fluids obeying an equation of state (EoS) of the type $p=p(e)$ 
(barotropic EoS), where $e=\rho(1+\epsilon)$ is the total mass-energy density 
($\rho$ and $\epsilon$ being, respectively, the rest-mass density and the 
internal specific energy), are a particular class of fluids which become 
familiar, for example, to cosmologists interested in the analysis of linear 
perturbations of radiation-dominated universes (see, e.g.,~\cite{Peacok99}), or
to astrophysicists interested in the physics of dense matter where the 
asymptotic EoS for high density and cold matter is barotropic (see, 
e.g.,~\cite{Camenzind07}). The mathematical properties of relativistic Euler 
equations for barotropic EoS have been studied in~\cite{ST93}. A particular 
case of a barotropic EoS is the {\it ultrarelativistic} EoS, where pressure is 
directly proportional to the energy density. Recently, several authors have 
paid particular attention to this ultrarelativistic EoS in order to find the 
exact solution of the relativistic hydrodynamical Riemann problem with nonzero 
tangential velocities~\cite{Mach10}, the corrugation instabilities occurring 
for solutions of that Riemann problem~\cite{Mach11}, or just in order to 
compare efficiencies of relativistic hydro-codes specifically designed for such
particular EoS~\cite{Ryu11}.

Twenty years ago, authors in~\cite{MIM91} paved the way for modern numerical 
relativistic hydrodynamics (RHD) extending to this field the strategy of using 
high-resolution shock-capturing (HRSC) methods. By exploiting the hyperbolic 
and conservative characters of the RHD equations, these authors (and 
collaborators) derived the full spectral decomposition of the Jacobian matrices
associated with the fluxes in a series of papers: \cite{MIM91}, in 1D 
(one-dimensional) general relativity hydrodynamics (GRHD); \cite{Font94} and 
\cite{Donat98}, in multidimensional special relativity hydrodynamics (SRHD); 
\cite{Banyuls97} (where the eigenvectors were given for a diagonal metric)
and \cite{Ibanez01} (where authors derived the full spectral decomposition:
right and left eigenvectors) in multidimensional GRHD. 
This numerical strategy has proved to be very fruitful in the field of 
computational RHD (see, e.g., the reviews~\cite{MM03} and \cite{Font08}). 

All the above papers (and works related with them) as well as all the current 
numerical codes based on these references make use of a general EoS of the form
$p=p(\rho,\epsilon)$. This general EoS contains some particular cases (see, 
e.g.,~\cite{LeV92}, \cite{Toro09}, ~\cite{Anile89}): (i) dust $(p=0)$. For this
EoS, motion equations can be reduced to the inviscid Burger's equation (in 1D).
(ii) A gamma-law EoS $(p=K\rho^{\gamma})$. This case makes one of the motion
equations (the energy equation) redundant.

This note is addressed to those researchers, working in relativistic 
astrophysics or numerical relativity, that might be interested in studying the 
evolution of barotropic fluids by using relativistic hydro-codes originally 
designed to use Godunov-type methods and a general EoS, of the type 
$p=p(\rho,\epsilon)$.

The evolution of a relativistic fluid is described by a system of equations 
which are the expression of local conservation laws: the local conservation of 
baryon number and energy-momentum, respectively,
\begin{eqnarray}
\label{J} \nabla_{\mu}J^{\mu} = 0, \\
\nabla_{\mu}T^{\mu\nu} = 0, \label{T}
\end{eqnarray}
the current $J^{\mu}$ and the energy-momentum tensor $T^{\mu\nu}$ being              
\begin{eqnarray}                                                                
J^{\mu} = \rho u^{\mu}, \\
T_{\mu\nu} = \rho h u_{\mu}u_{\nu} + p g_{\mu \nu}.                              
\end{eqnarray}
(Greek and Latin indices run, respectively, from 0 to 3 and from 1 to 3. In the
units we use, the speed of light is equal to one).

In the above equations, $h$ is the specific enthalpy, defined by 
$h = 1 + \varepsilon + p/\rho$, $u^{\mu}$ is the 4-velocity of the fluid, and 
$g_{\mu \nu}$ defines the metric of a general spacetime $\cal M$ where the fluid 
evolves, $\nabla_{\mu}$ stands for the covariant derivative.                                            

In the following we will restrict our analysis to the Minkowski spacetime. The 
above system of equations, (\ref{J}) and (\ref{T}), can be written as:                                
\begin{equation}
\frac{\partial{\bf F}^0({\bf w})}{\partial \,t} 
+ \frac{\partial{\bf F}^i({\bf w})}{\partial \, x^i} =
\frac{\partial{\bf F}^0({\bf w})}{\partial \,t} 
+ B^i \,\frac{\partial{\bf F}^0({\bf w})}{\partial \, x^i} = 0,
\label{F}                                                                       
\end{equation}                                                                      
where 
\begin{equation}
B^i \,=\, \frac{\partial {\bf F}^i({\bf w})}{\partial {\bf F}^0({\bf w})} 
\label{Bj}                                                                       
\end{equation}
are three $5 \times 5$ Jacobian matrices and 
${\bf w} = (\rho, v^{j}, \varepsilon)$ is the 5-vector of primitive variables. 
The components of the 3-velocity, $v^j$, are defined according to 
$\displaystyle v^j= \frac{u^j}{u^0}$, and the Lorentz factor, defined by 
$W \equiv u^0$, satisfies the familiar relation $W=(1-{\rm v}^{2})^{-1/2}$, 
where ${\rm v}^{2}= \eta_{ij} v^i v^j$, with $\eta^{ij}$ being the spatial 
metric of Minkowski spacetime.

The components of the 5-vector of conserved variables,
\begin{equation} 
{\bf F}^{0}({\bf w})  = (D,\,\, S^j,\,\, \tau) =  
	\left(\rho W,\,\, \rho h W^2 v^j,\,\, \rho h W^{2} - p - \rho W\right),  
\end{equation}
are, respectively, the rest-mass density, the relativistic momentum density and
the total energy density without the rest-mass density.

The three 5-vectors of fluxes are:
\begin{equation} 
{\bf F}^{i}({\bf w}) = \left(\rho W v^{i}, \,\,\rho h W^2 v^j v^i + 
p \, \eta^{ij},\,\,\rho h W^2 v^i - \rho W v^{i}   \right).
\end{equation}
Assuming an EoS of the form $p=p(\rho,\varepsilon)$, the local sound velocity 
$c_{s}$ is given by
\begin{equation}
h c_{s}^{2} = \chi + \left(\frac{p}{\rho^{2}}\right)\kappa  \,\,\,\,,\,\,\,\,
\chi = \frac{\partial p}{\partial \rho} \,\,\,\,,\,\,\,\,
\kappa = \frac{\partial p}{\partial \epsilon}.
\label{csRel}
\end{equation}
Authors in~\cite{Donat98} showed the characteristic structure of the 
multidimensional RHD system of equations. Some components of one 
righteigenvector, and also the determinant of the matrix of the 
righteigenvectors (Eqs.~(17) and (21), respectively, in~\cite{Donat98}) contain
the expression  ${\cal K} := \displaystyle{\frac{\tk}{\tk -c_s^2}}$, with 
$\tk := \displaystyle{\frac{1}{\rho}\,\,\frac{\partial p}{\partial \epsilon}}$. 
It is clear that ${\cal K}$ is not defined when $\tk-c_s^2 = 0$, and therefore,
in this case, the system of righteigenvectors given in~\cite{Donat98} is not 
well behaved. Regarding this point, the following proposition can be proved.

\noindent 
{\bf Proposition} {\it The thermodynamic function $\tk-c_s^2$ is identically 
zero iff $p=p_o(e), \, e:=\rho(1+\epsilon)$, with $p_o$ being an arbitrary 
function.}\\
Proof. 

$\tk - c_s^2 = 0$, taking into account the definition of $c_s^2$, can be shown 
easily to be equivalent to $(1+\epsilon) \, \tk - \chi = 0$. Introducing the 
auxiliary variables $t:=\log(1+\epsilon)$ and $x:=\log\rho$, we conclude that
$\displaystyle \tk - c_s^2 = 0 \Leftrightarrow \frac{\partial p}{\partial t} 
- \frac{\partial p}{\partial x} = 0$. The general solution of this 
advective-like partial differential equation (PDE) is $p = p_o(x+t)$, with 
$x+t = \log(\rho(1+ \epsilon)) = \log e$ and being $p_o$ an arbitrary function.

This conclusion can be also obtained in terms of functionals. 
$\tk - c_s^2 = 0$, from the definitions of $c_s^2$ and $e$, is equivalent to 
$\displaystyle \frac{\partial e}{\partial\epsilon}\frac{\partial p}{\partial\rho}
-\frac{\partial e}{\partial\rho}\frac{\partial p}{\partial\epsilon}=0$, which 
describes a functional dependence of $p$ and $e$, $p=p_o(e)$.

Hence, we have showed that $\tk - c_s^2 = 0 \Leftrightarrow p = p_o(e)$.
$\square$

From the above proposition, we conclude that the characteristic eigenvectors of
the SRHD equations~(\ref{F}) derived in~\cite{Donat98} have to be reviewed just
in the case of barotropic EoS. 

For the sake of conciseness, let us concentrate on the 1D version of the 
analysis presented in~\cite{Donat98}. The expressions for the righteigenvectors
are:
\begin{eqnarray}
{\bf r}_{0} =
\left(\frac{{\cal K}}{h W}, v, 1-\frac{{\cal K}}{h W}  \right), \\
{\bf r}_{\pm} = \left(1, h W (v{\pm}c_s), h W (1 {\pm} vc_s) - 1 \right).
\end{eqnarray}
Multiplying ${\bf r}_{0}$ by the normalization factor $\tk - c_s^2$, the 
determinant of the matrix of renormalized righteigenvectors, $\Delta_{1D}$, 
satisfies
\begin{equation}
\Delta_{1D} = \frac{2\, c_s^3\,h }{W}.
\label{det1}
\end{equation}
The singularity has been removed, both in the eigenvectors and the determinant,
and the system of eigenvectors is complete for a general EoS, including a 
barotropic one.

Analogously, in the multidimensional case, it can be showed that, given an 
arbitrary spacelike 3-vector $\zeta_j$, the general eigenvalue problem for the 
matrix $\zeta_j B^j$ (see equation~(\ref{F})) has the following renormalized 
righteigenvectors:
\begin{eqnarray}
{\bf r}_{01} =
\left( \frac{\tk}{h W}, (\tk-c_s^2)v^i, (\tk-c_s^2) - \frac{\tk}{h W} \right), \\
{\bf r}_{02,03} =
\left( W\,v_j\bar{\zeta}^j, h (\bar{\zeta}^i + 2 W^2 v_j \bar{\zeta}^j v^i), 
(2 h W - 1) W\,v_j\bar{\zeta}^j \right), \\
{\bf r}_{\pm} = \left(1, h W \left( \frac{({\cal A}_{\pm}-1)}{v_j \zeta^j}\zeta^i + v^i \right),
h W {\cal A}_{\pm} - 1 \right),
\end{eqnarray}
where $\displaystyle {\cal A}_{\pm}=\frac{(v_j \zeta^j)^2 - \zeta_j \zeta^j}
{\lambda_{\pm} v_j \zeta^j - \zeta_j \zeta^j}$, 
and $\bar{\zeta}_j$ denotes the two unitary spacelike 3-vectors which together 
with $\frac{\zeta_j}{\sqrt{\zeta_j\zeta^j}}$ form an orthonormal basis. The 
particular values of the above righteigenvectors in the direction 
$\zeta_j =(1,0,0)$ coincide, as it should be, with the ones derived 
in~\cite{Donat98}. The singularity has been removed and all the components of 
the eigenvectors are well defined. The determinant of the righteigenvectors, 
$\Delta$, does not vanish,
\begin{equation}
\Delta = \displaystyle{\frac{-c_s^2 W h^3 (\lambda_+ - \lambda_-) 
((v_j \zeta^j)^2 - \zeta_j \zeta^j)^2}
{(\lambda_+ \, v_j \zeta^j - \zeta_j \zeta^j)
(\lambda_- \, v_j \zeta^j - \zeta_j \zeta^j)\sqrt{\zeta_j \zeta^j}} \neq 0,}
\end{equation}
and the system of eigenvectors is complete (even for a barotropic EoS).

Therefore, the main conclusion is that, by an appropriate renormalization of 
one of the eigenvectors derived in~\cite{Donat98}, it is possible to use the 
corresponding numerical hydro-code for a general EoS, including the case of 
barotropic EoS. Similar renormalization is required in RMHD when the 
characteristic eigenvalues of the system of equations become 
degenerate~\cite{Anton10}.   

A second strategy is possible: to rewrite down the SRHD equations for a 
barotropic EoS. It is the one proposed in~\cite{Mach10, Mach11, Ryu11}. Yes, 
indeed, quoting Anile in~\cite{Anile89} (page 11): `Since the rest-mass 
density does not intervene in the state equation, these fluids can be described
solely by the conservation of energy and momentum equations. The mass 
conservation equation decouples from the others and can be solved after the 
fluid motion has been determined'. The above Anile's quotation defines the 
(alternative) computational strategy to follow. 

The conservation of momentum and energy equations governing the evolution of a 
barotropic fluid, i.e. the one satisfying a barotropic causal EoS, are
\begin{eqnarray}
\label{eqsSRHD1}
\partial_t [(e+p) W^2 v^i] + \partial_j [(e+p) W^2 v^i v^j + p \eta^{ij}] = 0, \\
\label{eqsSRHD2}
\partial_t [(e+p) W^2 - p] + \partial_j [(e+p) W^2 v^j] = 0.
\end{eqnarray}
The new vector of conserved variables is $U = ((e+p) W^2 v^i, (e+p) W^2 - p)$. 
The system of equations~(\ref{eqsSRHD1}) and (\ref{eqsSRHD2}) can be cast into 
the form of a quasi-linear system of PDEs:
\begin{equation}
\label{e:evol-var}
	\partial_t U + A^j \partial_j U = 0, \,\,\,\, A^j = \left(
	\begin{tabular}{c|c}			
	$A^j_1$ & $A^j_2$ \\ \hline
	$\delta^j_k$ & 0
	\end{tabular} \right),
\end{equation}
where $A^j_1 = v^j \, \delta^i_k + [(1 - v^2 c_s^2) \, \delta^j_k \, v^i 
+ 2 \, c_s^2 \, v^j \, v_ k \, v^i - 2 \, c_s^2 \, v_k \, \eta^{ij} \,] \,
(1 - v^2 c_s^2)^{-1}$ and $A^j_2 = [- (1 + c_s^2) \, v^j \, v^i 
+ c_s^2 \, (1 + v^2) \, \eta^{ij} \,](1 - v^2 c_s^2)^{-1}$.

Let $\zeta_j$ be an arbitrary spacelike vector. The eigenvalues of the matrix
$\zeta_j A^j$ are
\begin{eqnarray}
	\lambda_0 = \zeta_j v^j, \\
	\lambda_\pm = \frac{(1-c_s^2) \zeta_j v^j \pm c_s (1-v^2)^{1/2} 
((c_s^2 - 1) (\zeta_j v^j)^2 + (1 - v^2 c_s^2)\zeta_j \zeta^j)^{1/2}}
{1 - v^2 c_s^2},
\end{eqnarray}
where $\lambda_0$ has multiplicity 2. The corresponding eigenvalues associated 
with previous set of eigenvalues are
\begin{eqnarray}
	r_0 = \left(\begin{tabular}{c}
						  $(1 - v^2) \bar{\zeta}^i + 2 \, \bar{\zeta}_j v^j \, v^i$ \\
						  $2 \, \bar{\zeta}_j v^j$
	            \end{tabular}\right), \\
	r_\pm = \left(\begin{tabular}{c}
						  $(\zeta_j v^j - \lambda_\pm) \zeta^i 
					      + (\lambda_\pm \, \zeta_j v^j - \zeta_j \zeta^j) v^i$ \\
						  $(\zeta_j v^j)^2 - \zeta_j \zeta^j$
	            \end{tabular}\right),
\end{eqnarray}
where $\bar{\zeta}_j$ denotes, as previously, the two unitary spacelike 
3-vectors which together with $\frac{\zeta_j}{\sqrt{\zeta_j\zeta^j}}$ form an 
orthonormal basis. The system of eigenvalues form a complete basis. Therefore, 
the system of equations~(\ref{e:evol-var}) is hyperbolic.

In brief, according to this second strategy and for a barotropic EoS, the 
numerical procedure, regarding the use of Godunov-type methods, would consist 
in solving, first, the system~(\ref{e:evol-var}). The rest-mass density can be 
obtained, in a second step, by solving the continuity equation in the field 
velocity provided by the first step. Quoting Anile in~\cite{Anile89} (page 11):
`...the only role of the continuity equation is to provide a volume tracer.'

\vspace*{0.5cm}

\noindent {\bf Summary}

The characteristic formulation of the RHD equations carried out 
in~\cite{Donat98} needs a minor modification for a barotropic EoS in order to 
remove the singularity in one of the eigenvectors.

Any user of a relativistic hydro-code, based on Godunov-type techniques and 
designed for a general EoS $p=p(\rho, \epsilon)$, that might be interested in 
using a barotropic EoS, can follow one of the following two strategies.
(A) Just to renormalize one of the eigenvectors, such as we have shown above. 
(B) To rewrite explicitly the SRHD equations for a barotropic EoS.

In terms of efficiency, it is obvious that strategy (B) is the most 
recommendable, since there is one equation less in the system. However, we 
recommend strategy (A) due to the fact that it can be applied in studies of
flows obeying a general EoS, including the barotropic ones. Moreover, strategy 
(A) allows one to handle with those flows that, during their evolution, might 
change their thermodynamical properties in such a way that the EoS admits a 
barotropic branch as, e.g., the asymptotic behavior of the EoS for dense matter
at high densities. In such a case, strategy (A) would be the only one suitable 
for a global description of the evolution of these flows.

\ack Work partially supported by the Spanish Ministry of Science 
(grants: AYA2010-21097-C03-01 and AYA2010-21097-C03-2).
I. C.-C. acknowledges support from Alexander von Humboldt Foundation.

\end{document}